\documentclass{PoS}
\usepackage{multirow}
\usepackage{sidecap}

\title{Two-nucleon scattering in multiple partial waves}

\ShortTitle{Two-nucleon scattering in multiple partial waves}

\author{\speaker{Amy Nicholson}%
         \thanks{LLNL-PROC-678486}\\
        Department of Physics, University of California, Berkeley\\
        E-mail: \email{anicholson@berkeley.edu}}

\author{Evan Berkowitz, Enrico Rinaldi, Pavlos Vranas\\
Physics Division, Lawrence Livermore National Laboratory\\
        E-mail: \email{berkowitz2@llnl.gov}, \email{rinaldi2@llnl.gov}, \email{vranas2@llnl.gov}}
        
\author{Thorsten Kurth\\
Nuclear Science Division, Lawrence Berkeley National Laboratory\\
Department of Physics, University of California, Berkeley\\
        E-mail: \email{tkurth@lbl.gov}}
        
\author{B\'{a}lint~Jo\'{o}\\
Theory Center, Jefferson National Accelerator Facility\\
        E-mail: \email{bjoo@jlab.org}}  
        
\author{Mark Strother \\%
        Department of Physics, University of California, Berkeley\\
        E-mail: \email{mcstro@berkeley.edu}}

\author{Andr\'{e} Walker-Loud\\
Nuclear Science Division, Lawrence Berkeley National Laboratory\\
Theory Center, Jefferson National Accelerator Facility\\
Department of Physics, The College of William \& Mary\\
        E-mail: \email{awalker-loud@lbl.gov}}  

\abstract{We determine scattering phase shifts for $S,P,D$, and $F$ partial wave channels in two-nucleon systems using lattice QCD methods. We use a generalization of L\"uscher's finite volume method to determine infinite volume phase shifts from a set of finite volume ground- and excited-state energy levels on two volumes, $V=(3.4 \, \mathrm{fm})^3$ and $V=(4.5\,  \mathrm{fm})^3$. The calculations are performed in the SU(3)-flavor limit, corresponding to a pion mass of approximately 800 MeV. From the energy dependence of the phase shifts we are able to extract scattering parameters corresponding to an effective range expansion.}

\FullConference{The 33rd International Symposium on Lattice Field Theory\\
                 14 -18 July  2015\\
                 Kobe International Conference Center, Kobe, Japan}

\begin{document}

\section{Introduction}
Lattice QCD provides a quantitative tool with controllable systematics for understanding low-energy nuclear physics from first principles. While the two-nucleon interactions are well-known from experiment, three- and higher-body interactions are still poorly constrained, and the lattice, in principle, gives us a unique opportunity to extract these directly from theory. However, its application to few- and many-nucleon systems remains a formidable challenge. In order to make progress toward this goal we must first have complete control over the two-nucleon system for two main purposes. The first is to have the ability to disentangle two-body contributions from the pure three- and higher-body interactions. The second is to develop techniques for extracting signals from lattice QCD for the desired states and resolving such issues as how to construct good operators, how to distinguish the closely spaced energy levels that arise in multi-particle states, and how to interpret finite volume results in terms of infinite volume observables.

Another motivation for studying two-nucleon interactions using lattice QCD is that we are able to tune the Standard Model parameters as we like in order to explore the dependence of observables on quantities such as the quark mass. It is well known that nucleon interactions are finely tuned in certain channels, but it is not clear how strongly this fine-tuning depends on the underlying parameters. Varying quark masses also gives us another handle for matching coefficients in effective field theories, which may then be used to predict less well-known quantities. Finally, in order to calculate two-nucleon matrix elements on the lattice, for example, the parity-violating neutral weak current, the two-nucleon scattering phase shifts and their derivatives must be determined to convert finite volume results to the desired infinite volume matrix element. 

In these proceedings we present results for $S$-, $P$-, $D$-, and $F$-wave scattering phase shifts determined from finite-volume spectra calculated using lattice QCD. The finite-volume energies are related to infinite-volume phase shifts via a generalization of the well-known L\"uscher relation. Most of these results have appeared previously in Ref.~\cite{Berkowitz:2015eaa}.

\section{Correlation functions}

To calculate the finite-volume spectra necessary for extracting the phase shifts we must form correlation functions using operators that have good overlap with the states of interest. Our correlation functions are composed using displaced nucleon operators at the source, which are not only necessary for probing negative parity states, but also possess superior overlap onto the low-energy even parity states. These are formed by choosing a good interpolating operator for a single nucleon, $N^\dagger(t_0,\mathbf{x}_0 )$, at Euclidean time $t_0$ and spatial position $\mathbf{x}_0$. A second nucleon is displaced along the vector $\Delta \mathbf{x}$, $N^\dagger(t_0,\mathbf{x}_0 + \Delta \mathbf{x})$. The direction of the displacement corresponds to one of three configurations: ``face'', with $\Delta \mathbf{x} \propto \mathcal{R} (1,0,0)$, where $\mathcal{R}$ represents any rotation or reflection in the cubic group and $(x,y,z)$ is a 3-D vector, ``edge'', with $\Delta \mathbf{x} \propto \mathcal{R} (1,1,0)$, or ``corner'', with $\Delta \mathbf{x} \propto \mathcal{R} (1,1,1)$, so-named due to their locations on a cube surrounding $\mathbf{x}_0$. Due to their differing properties under cubic rotations, the three types of sources have considerably different overlaps onto the various cubic irreps. We choose to focus on ``corner'' and ``edge'' sources due to the lack of overlap of the ``face'' sources with several of the channels and shells of interest. We also vary the distance, $|\Delta \mathbf{x}|$, in order to maximize overlap with as many shells as possible. We find that for low-energy states large displacements, of order $\sim L/4$, are ideal.

The spins of these displaced nucleon-nucleon operators are combined into either total spin-0 or spin-1 states, and are then projected using spherical harmonics,
\begin{eqnarray}
\sum_{\mathcal{R}} Y_{\ell m}(\mathcal{R} \widehat{\Delta \mathbf{x}}) \bar{N}(\mathbf{x}_0+\mathcal{R}\Delta\mathbf{x})N(\mathbf{x}_0) \ .
\end{eqnarray}
The spins and angular momenta are then combined using the standard Clebsch-Gordan coefficients into the infinite volume channels of interest. In many cases these infinite volume angular momentum labels do not correspond to unique quantum numbers due to the reduced symmetry of the box. However, different angular momentum projections having overlap onto the same cubic irrep, such as the operators labeled $^3S_1$, $^3D_1$, and $^3D_3$, all having overlap onto the $T_1^+$ channel, may be used as additional handles for disentangling the various energy levels within a given channel. The final projection onto the cubic irreps is carried out using the subduction matrices presented in \cite{Dudek:2010wm}.

At the sink we form operators in momentum space corresponding to the exact non-interacting eigenstates of the box in the various channels \cite{Luu:2011ep}. This is done using the same projection method as was performed at the sink. In this case, however, one unit of displacement in momentum space corresponds to the non-interacting $n^2=1$ shell, and so on. The use of momentum space operators at the sink also allows us to project the system onto zero total momentum, eliminating boosted excited states and also removing any complications associated with the cubic symmetries as applied to boosted systems. We find that the use of operators corresponding to exact non-interacting eigenstates gives clear separation of the data between energy levels. In other words, there is very little mixing between the non-interacting shells once interactions are turned on, so we are able to extract several energy levels cleanly using this method.

\section{Spectrum}

For the lattice calculation we use the isotropic clover configurations generated by the JLab/W\&M group. The lattice spacing is $b\sim 0.145$ fm, and we use two volumes, $(24)^3\times 48$ and $(32)^3\times 48$. Due to the significant signal-to-noise problem in the two-nucleon sector, which is further amplified for higher partial waves, we use configurations corresponding to the SU(3) point, with pion and kaon masses of $\sim 800$ MeV. These are the same configurations used previously by NPLQCD for $S$-wave nucleon-nucleon scattering \cite{Beane:2013br}. 

\begin{figure*}[t]
\centering
\includegraphics[width=0.48\textwidth]{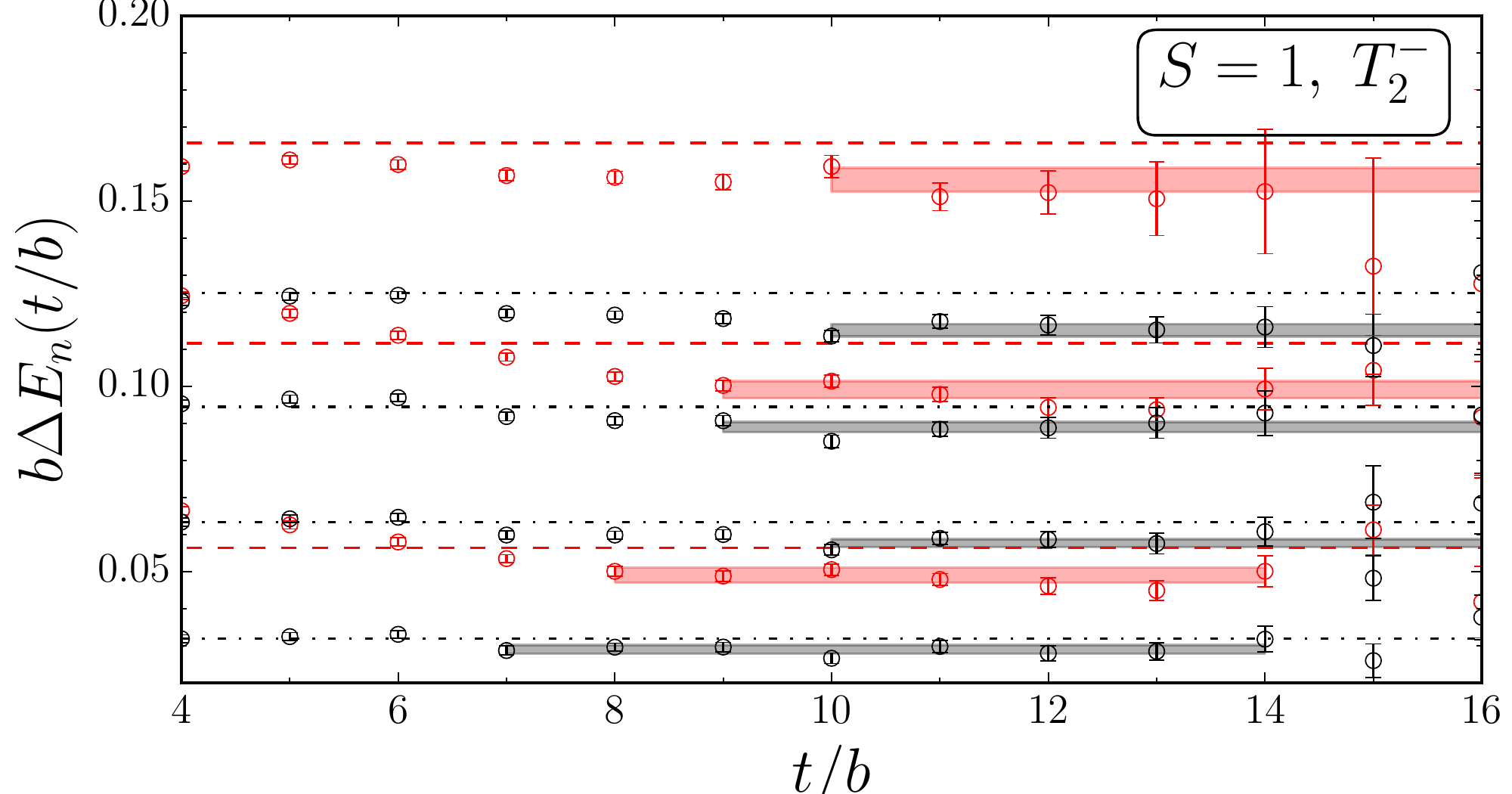}
\includegraphics[width=0.48\textwidth]{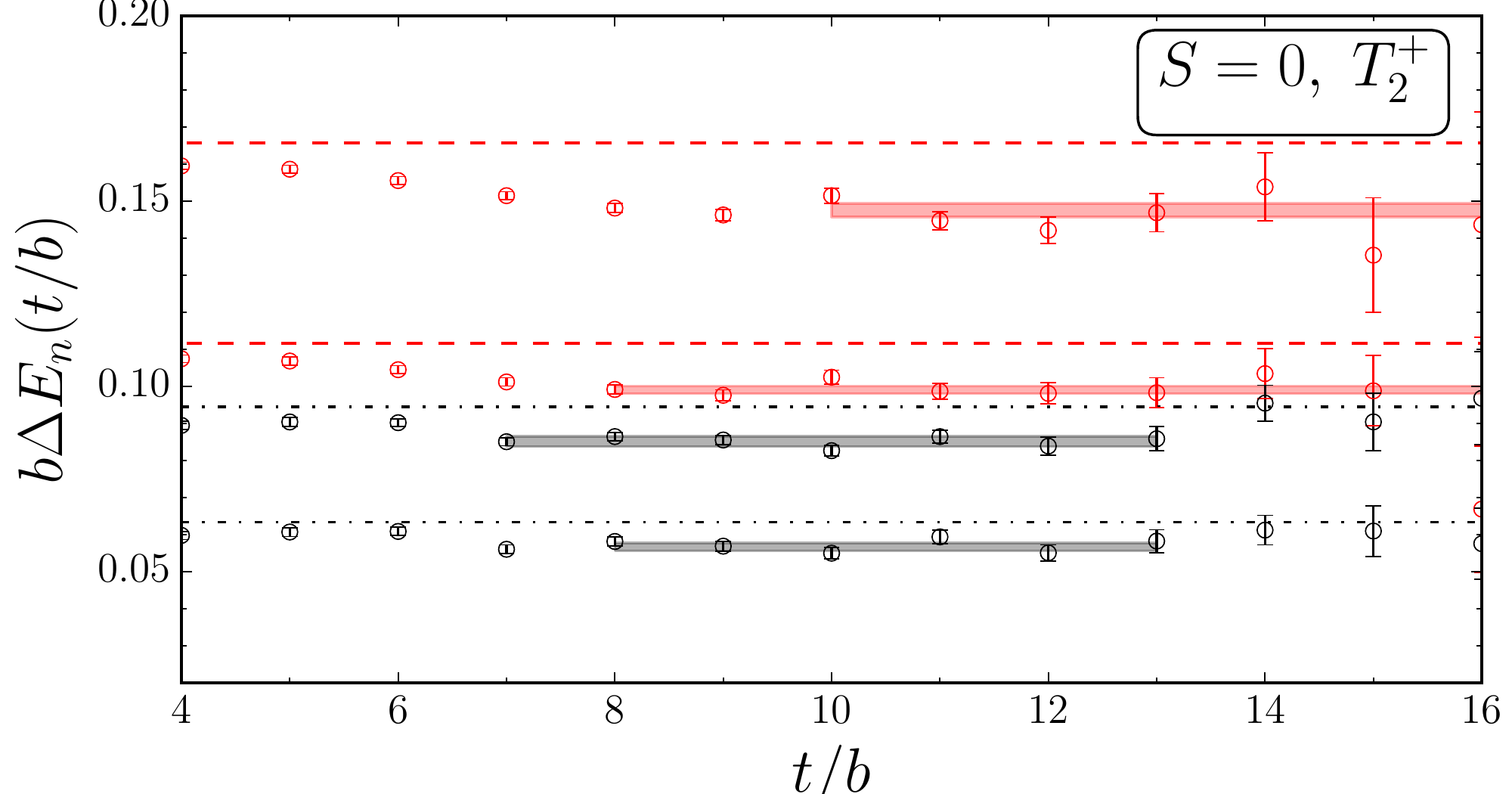}
\caption{\label{fig:effective_masses} Effective mass plots for the interaction energies, $\Delta E_n= 2\sqrt{m_N^2 +q_n^2}-2m_N$, in lattice units for the $T_2^-$ spin triplet and $T_2^+$ spin singlet channels. Multiple energy levels, corresponding to the non-interacting shell labeled by $n$, are shown, and red (black) points correspond to $L=24$ ($32$). Horizontal bands show fits to the data (see text). Dashed (dot-dashed) lines indicate the energy levels of the non-interacting systems.}
\end{figure*}

In Fig.~\ref{fig:effective_masses} we show some examples of effective mass plots for two different cubic irreps in two volumes. The dashed and dot-dashed lines represent the non-interacting energy levels corresponding to the operators used to generate each energy level. The bands are fits to the energies including statistical and fitting systematic errors. The projection onto non-interacting energy levels at the sink gives us clean separation between the interacting energy levels, and allows us to extract several states in each volume. We also have a resolution in our energies of a few sigma from the non-interacting energies in several cases. This is crucial because the non-interacting energies correspond to poles in the L\"uscher method, so uncertainties become drastically magnified near these regions when determining the phase shifts.

\section{Phase shifts}
In order to extract scattering phase shifts we use the L\"uscher method, which relates the energies of two particle systems in a periodic box to infinite volume phase shifts. In general, because partial waves mix in a periodic box due to the reduced symmetry, in addition to partial wave mixing in nature, such as the $S$- and $D$-wave mixing observed in the deuteron channel, the quantization condition is a matrix equation, $\det_{Jm_JS}\left[ \mathcal{M}^{-1}+ \delta \mathcal{G}^{V} \right]= 0$, where the elements extend over the relevant partial wave channels. Here $\mathcal{M}$ contains the infinite volume scattering amplitudes and $\delta \mathcal{G}$ is a kinematic function of the volume and the momenta. Solving such an equation requires several data points for different channels at the same energies. However, because we have only discrete sets of volumes (or potentially, boosts) this is not directly possible, so in general either the effective range expansion (ERE) or some modeling is required to interpolate between discrete data points in order to solve these equations.

Due to the substantially increased difficulty associated with solving the general matrix equation, in this work we neglect all partial wave mixing as a first step, using only the lowest partial wave coupling to a given cubic irrep. While higher partial waves are kinematically suppressed at low energies, it is unclear whether this will be the case at the energies we extract, and any effects of partial wave mixing should be explored in future work. Once this simplification is made the L\"uscher relations become simple. Ignoring partial waves with $\ell\geq4$, the spectra satisfy the quantization condition~\cite{Briceno:2013lba}
\begin{eqnarray}
\label{eq:1D_QCs}
\frac{q\cot\delta_\Lambda(q)}{4\pi}= c_{00}(q^{2})
	+\alpha_{4,\Lambda} \frac{c_{40}(q^{2})}{q^4}
	+\alpha_{6,\Lambda} \frac{c_{60}(q^{2})}{q^6},
\end{eqnarray}
where $q$ is the on-shell relative momentum of the two-particle system, $q^{2}=\frac{E^{2}_{NN}}{4}-m_{N}^2$, and is determined from the lattice spectrum, $\alpha_{\ell,\Lambda}$ are constants given in Ref.~\cite{Briceno:2013lba}, and the $c_{\ell m_\ell}$ are kinematic, non-linear functions that depend solely on the momentum and the volume 
\begin{eqnarray}\label{eq:clm}
c_{\ell m_\ell }(q^{2})
&=&\frac{\sqrt{4\pi}}{L^3}\left(\frac{2\pi}{L}\right)^{\ell -2}~\sum_{\mathbf r \in \mathbb{Z}^3}\frac{|\mathbf{r}|^\ell Y_{\ell m_\ell }(\mathbf{r})}{(r^2-q^2)}. \label{eq:clm}
\end{eqnarray} 
The scattering phase shift, $\delta_\Lambda(q)$, corresponds to the partial wave that primarily couples to the $\Lambda$ irrep. 

\begin{figure*}[t]
\centering
\includegraphics[width=0.42\textwidth]{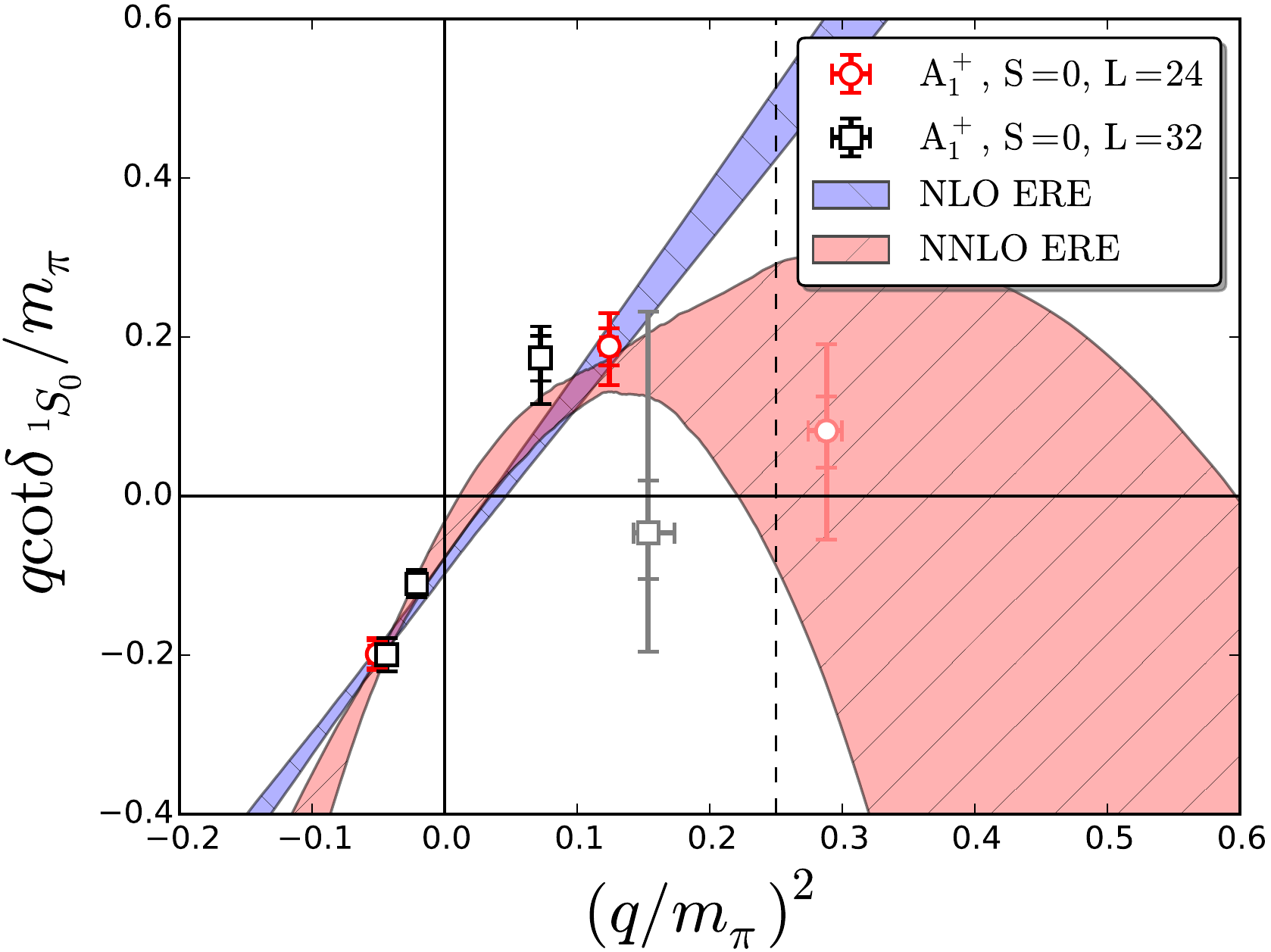}
\hspace{0.5cm} \includegraphics[width=0.42\textwidth]{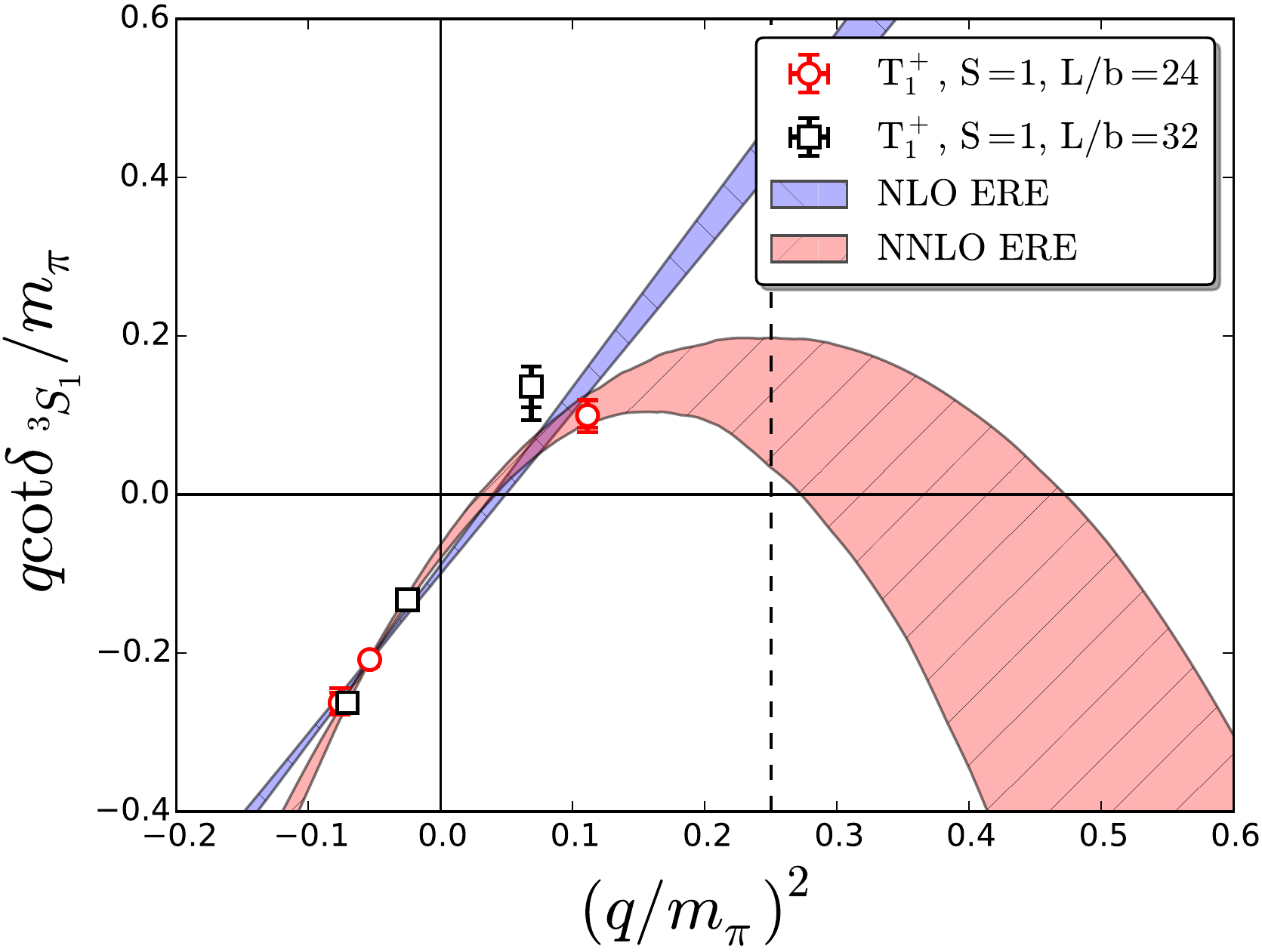}
\caption{\label{fig:swave}
$S$-wave scattering phase shifts and their corresponding ERE fits as a function of the lattice momenta, in units of the pion mass. The dashed vertical lines indicate the momentum at which the $t$-channel cut occurs. 
}
\end{figure*}

In Fig.~\ref{fig:swave}, we present results for $S$-wave scattering. The bands represent fits to the ERE, 
\begin{equation}
q^{2\ell+1}\cot\delta_\ell = -\frac{1}{a_\ell}+\frac{1}{2}r_\ell\,q^2
	+\frac{1}{4!}P_\ell q^{4}+\mathcal{O}(q^6),
\label{eq:ERE}
\end{equation} 
where $a_\ell$, $r_\ell$, and $P_\ell$ are the scattering length, effective range, and shape parameter for $\ell=0$ and the corresponding parameters of the ERE for $\ell>0$. The dashed vertical line represents the t-channel cut at $q=m_{\pi}/2$, above which the ERE is expected to break down. The L\"uscher formalism, on the other hand, holds for all energies below the $NN\pi$ threshold, which is well above the energies considered. 

These $S$-wave phase shifts have been determined previously on the same configurations by the NPLQCD collaboration \cite{Beane:2013br,Beane:2012vq} using local nucleon-nucleon interpolating fields. We find generally good agreement, including the identification of deeply bound states in both channels, giving us confidence in the use of our displaced operators. However, we find additional negative energy states near threshold to which the local operators do not couple strongly. This state may correspond to a second infinite volume bound state in the $^3S_1$ channel, which is also indicated by the location of a pole near threshold in the derived scattering amplitude. However, the state is very close to threshold and with our current precision we are not able to determine whether it is a true shallow bound state or a scattering state.

\begin{figure*}[t]
\centering
\includegraphics[width=0.42\textwidth]{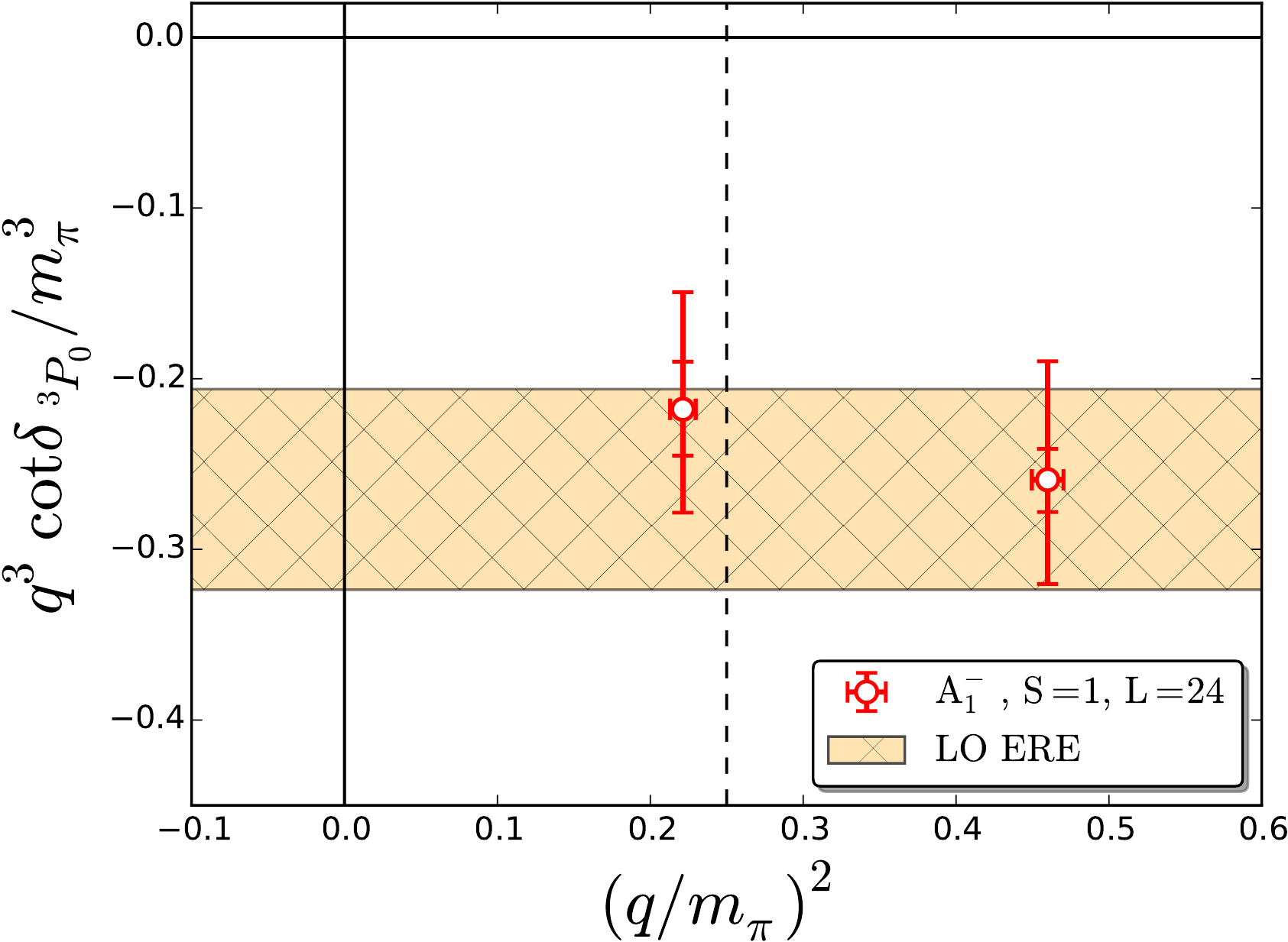}
\hspace{0.5cm} \includegraphics[width=0.42\textwidth]{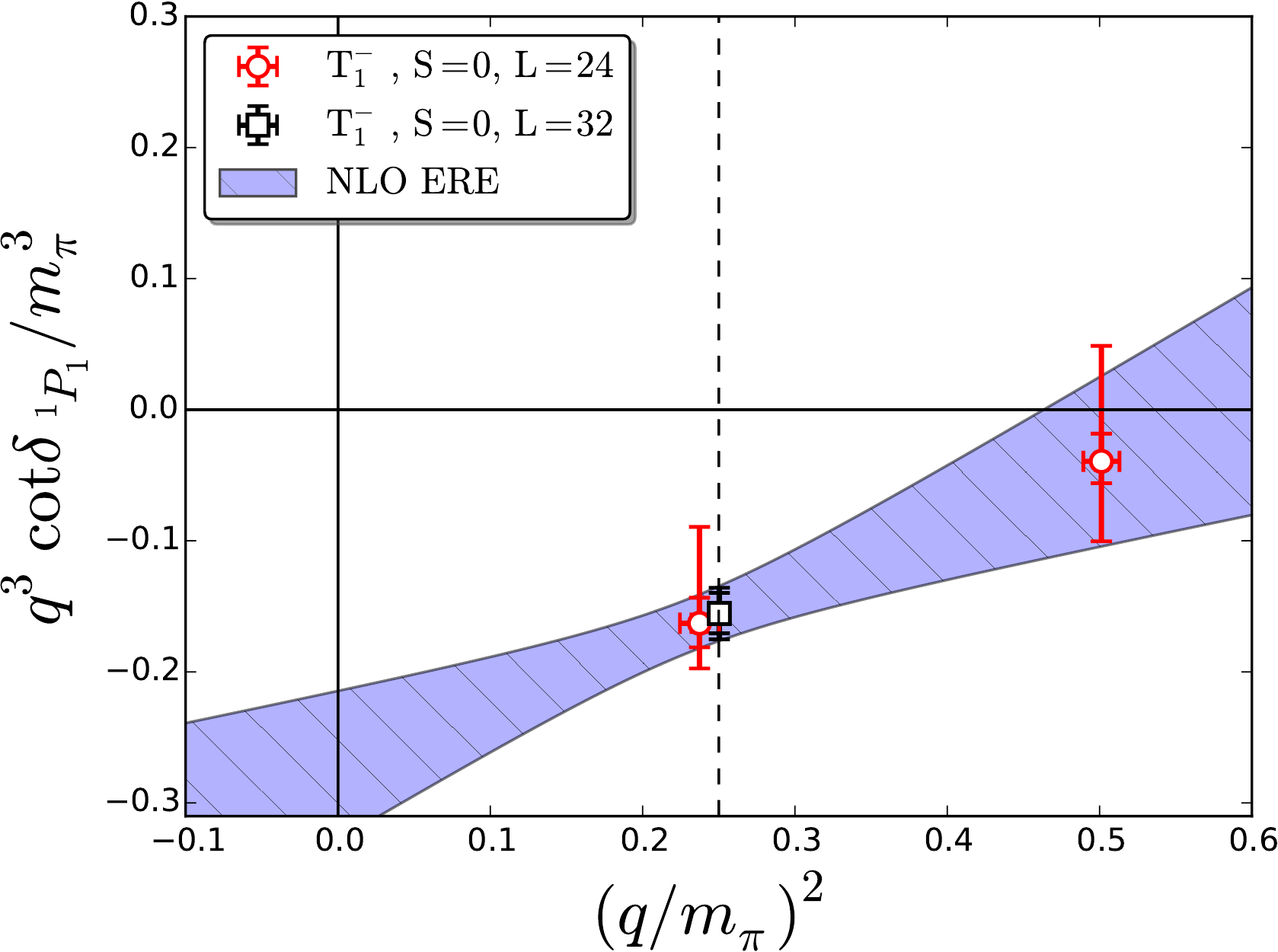} \\
\vspace{0.5cm}
\includegraphics[width=0.42\textwidth]{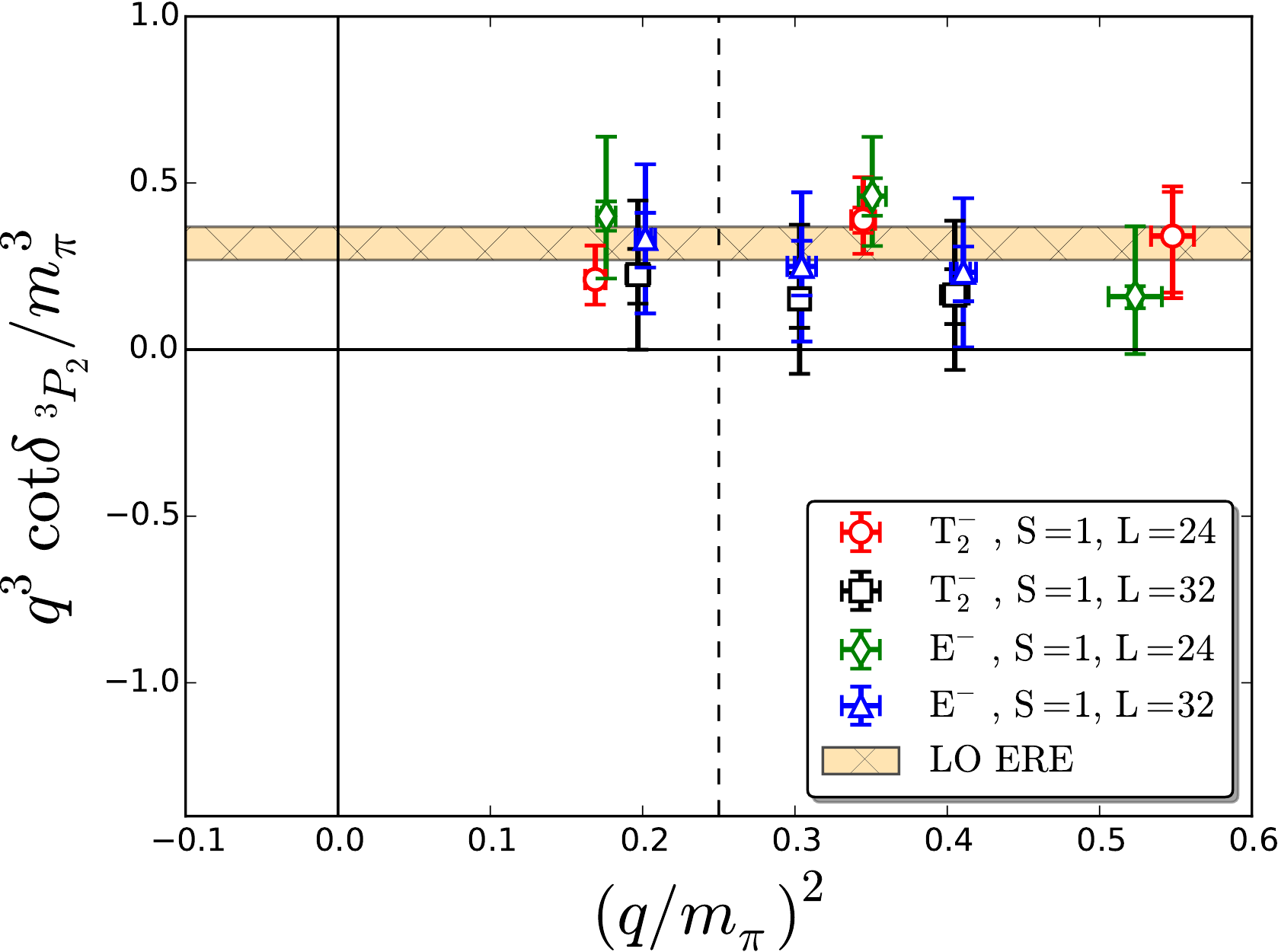} 
\hspace{0.5cm} \includegraphics[width=0.39\textwidth]{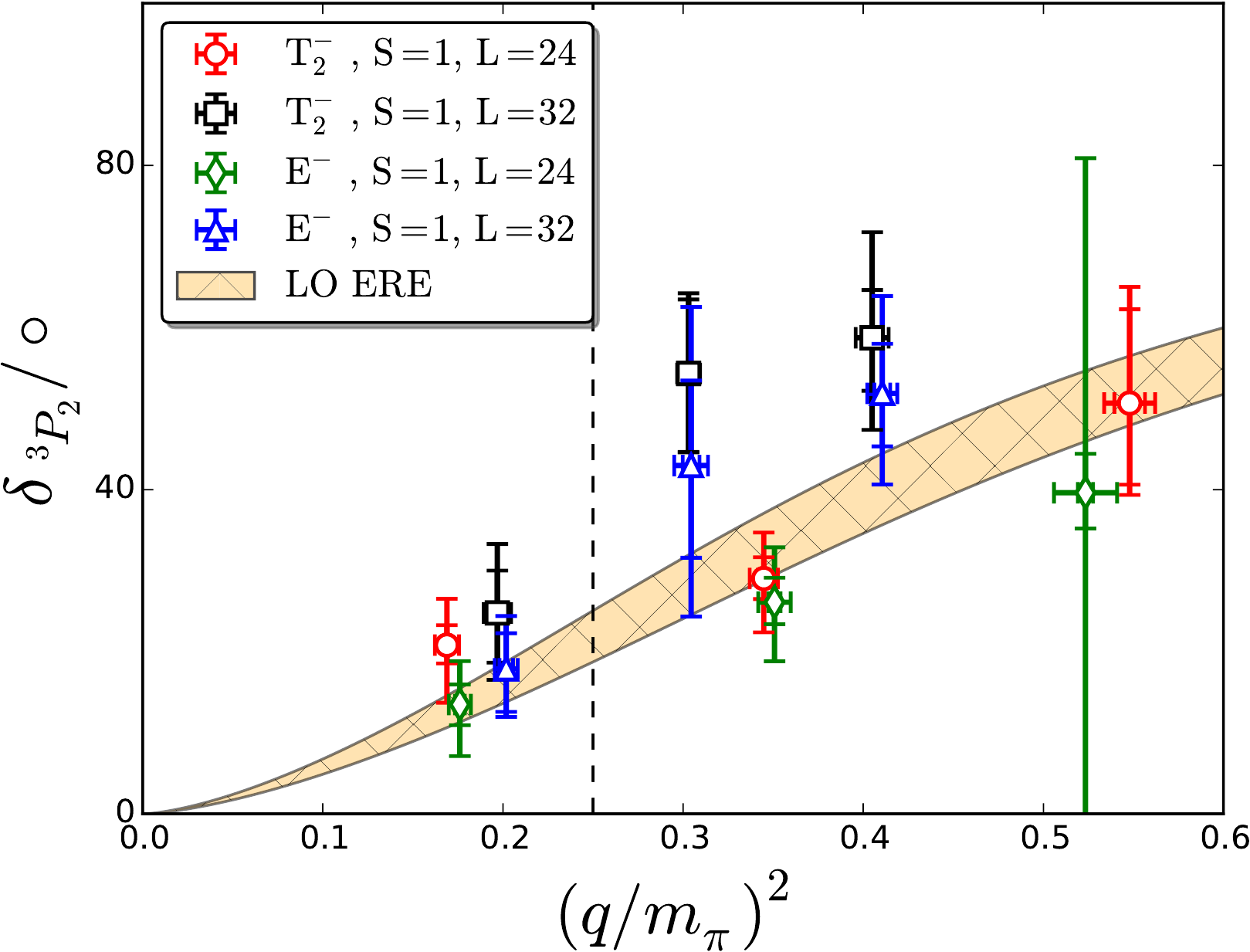}
\caption{\label{fig:pwave}
$P$-wave scattering phase shifts and their corresponding ERE fits as a function of the lattice momenta, in units of the pion mass. The dashed vertical lines indicate the momentum at which the $t$-channel cut occurs. In the lower right panel we also show the phase shift $\delta_{\,^3\!P_2}$ as a function of the lattice momenta.
}
\end{figure*}

Results for $P$-wave channels are shown in Fig.~\ref{fig:pwave}. The $^3P_0$ channel is of particular interest because it is necessary for calculating the $\Delta I=2$ hadronic parity violating matrix element involving the transition from $^3P_0$ to $^1S_0$ \cite{Kurth:2015}. The $^3P_2$ channel has contributions from multiple cubic irreps, and we find remarkably consistent behavior for the phase shift between the different irreps and volumes, illustrating the power of the operators and the finite volume formalism, despite the fact that we neglect partial wave mixing effects. Furthermore, we find no evidence of breakdown of the effective range expansion even above the t-channel cut. This is illustrated beautifully by looking directly at the phase shift in the lower right panel. The shape of the phase shift agrees qualitatively with experiment, even at $m_{\pi} \sim 800$ MeV, as well as calculations performed by HalQCD \cite{Murano:2013xxa}.

\begin{figure*}[t]
\centering
\includegraphics[width=0.42\textwidth]{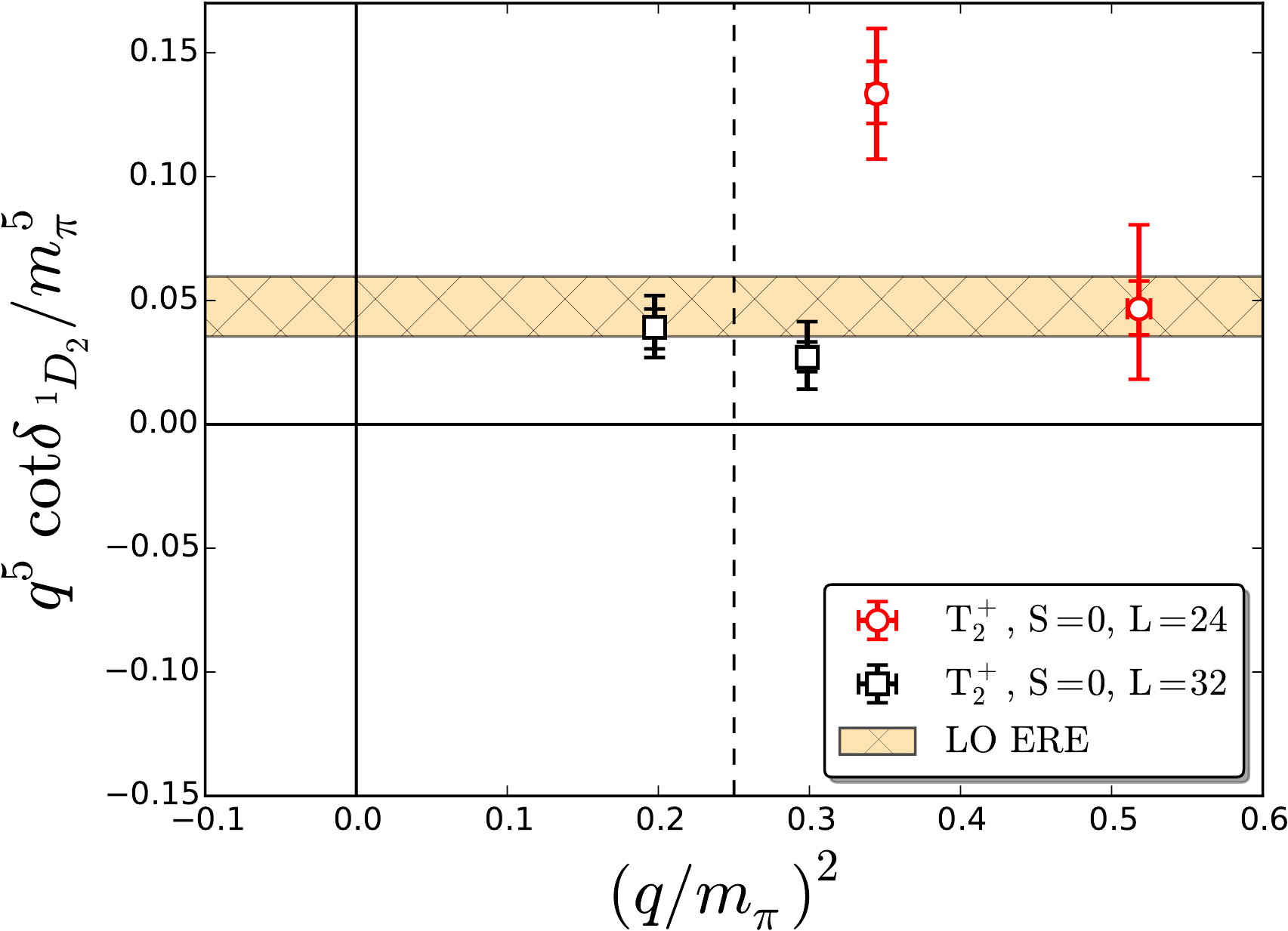}
\hspace{0.5cm} \includegraphics[width=0.42\textwidth]{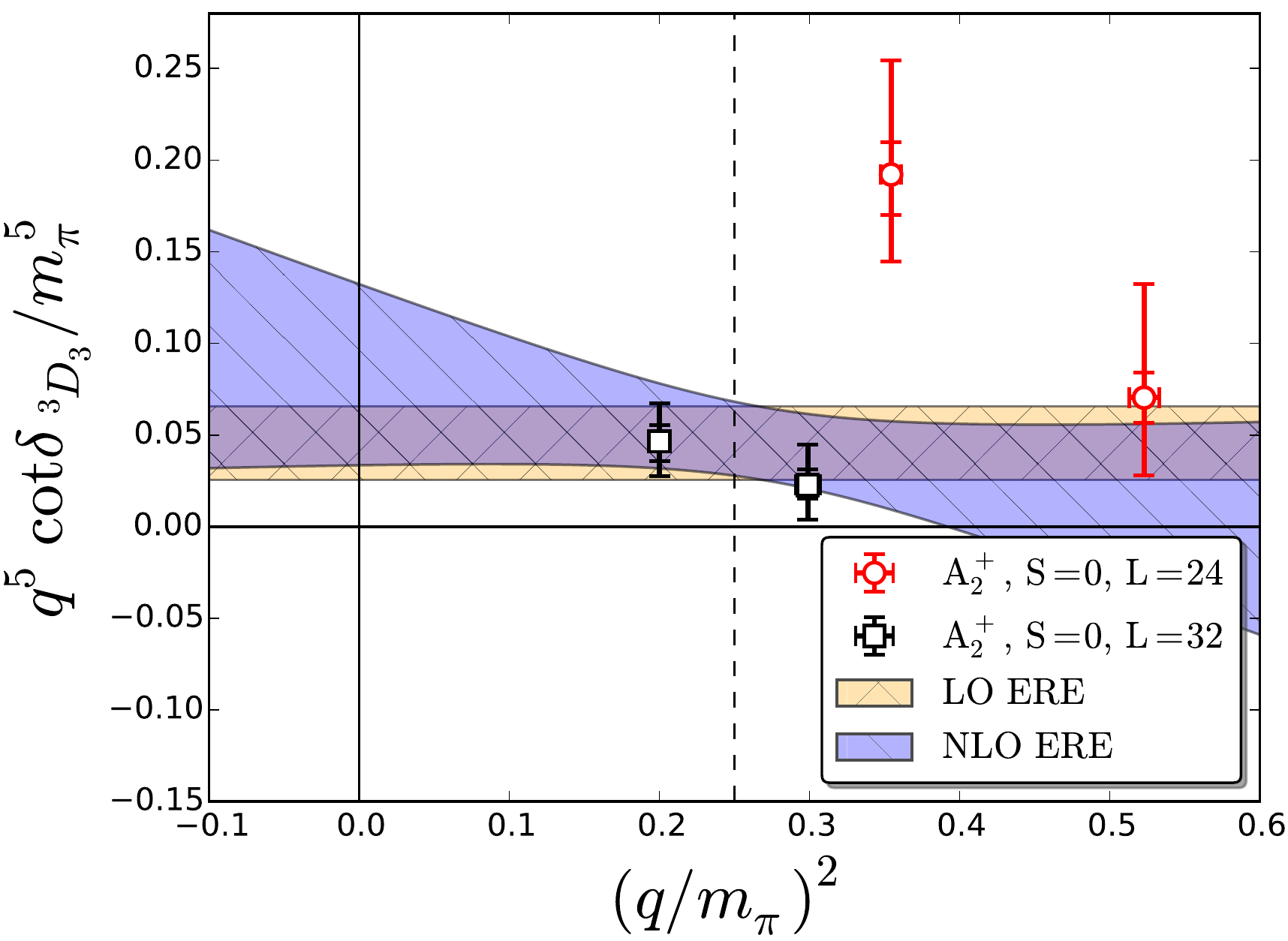} \\
\vspace{0.5cm}
\includegraphics[width=0.42\textwidth]{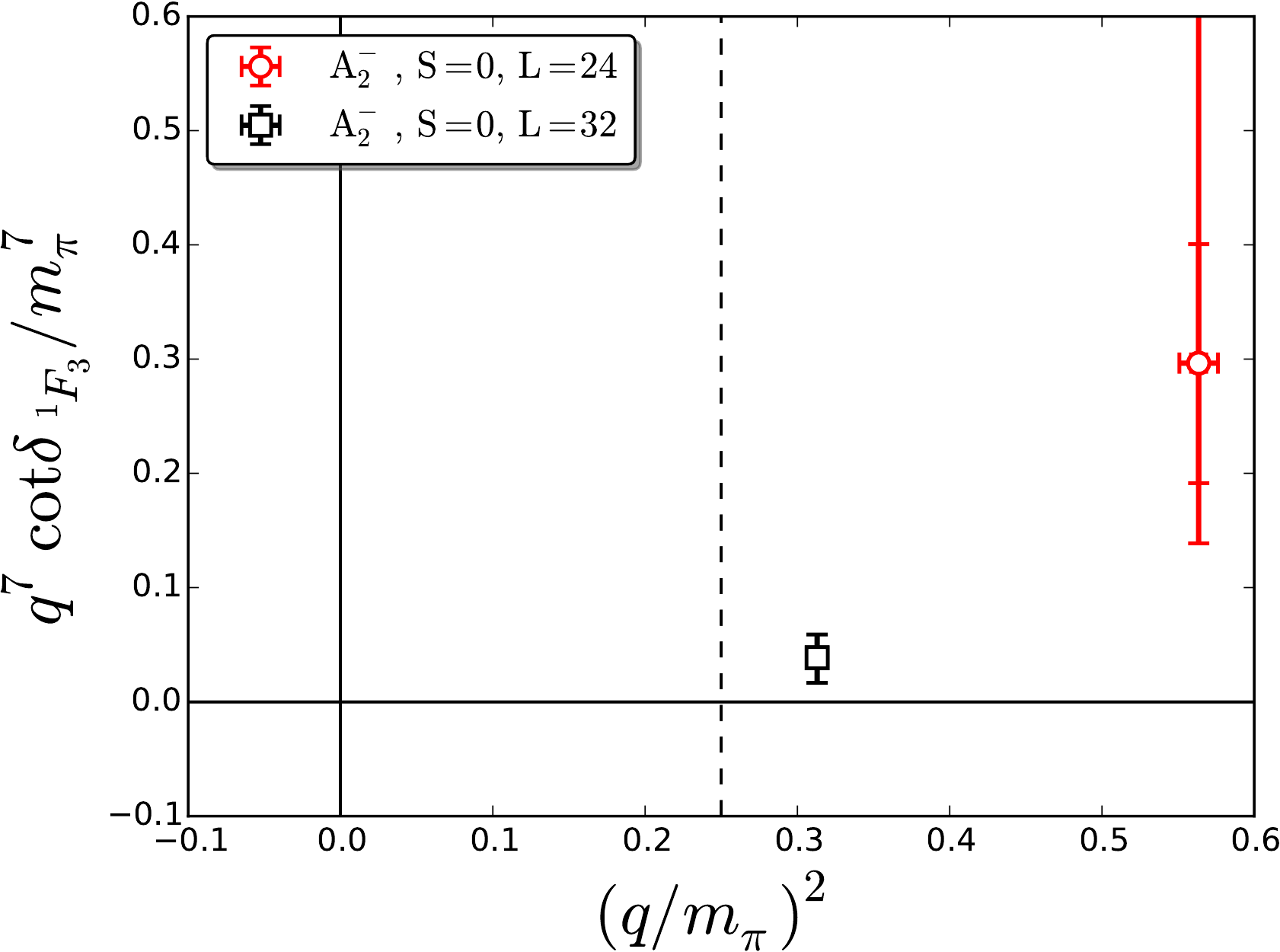}
\caption{\label{fig:dwave}
$D$-wave (upper) and $F$-wave (lower) scattering phase shifts and their corresponding ERE fits as a function of the lattice momenta, in units of the pion mass. The dashed vertical lines indicate the momentum at which the $t$-channel cut occurs.}
\end{figure*}

In Fig.~\ref{fig:dwave} we show results for $D$- and $F$-wave scattering. Because the lowest shells are insufficient to resolve higher partial waves, the number of low-energy states we are able to access becomes limited. However, we are still able to resolve non-zero signals in these channels. In particular, we find that the $F$-wave scattering phase shift is quite small even at these moderate energies, which may be an indication that partial wave mixing in $P$-wave channels is relatively small as well. However, as stated previously, to truly determine the size of these effects on $P$-wave scattering the $F$-wave contributions in the same channels must be disentangled using the full finite volume relations.

\section{Conclusions}

We have presented a first-principles calculation of nucleon-nucleon scattering in $S$-, $P$-, $D$-, and $F$-wave scattering channels. Our operator setup allows us to cleanly extract multiple energy levels for a given volume, and for the first time to present results for partial waves with $\ell >0$ using the L\"uscher finite volume method. We find deeply bound states in both the $^1S_0$ and $^3S_1$ channels, in agreement with Refs.~\cite{Orginos:2015aya,Beane:2013br, Yamazaki:2012hi,Yamazaki:2015asa}, in addition to a possible second bound state near threshold in the $^3S_1$ channel which was not previously found in calculations incorporating only local sources. We are able to extract both the parity even and odd states necessary for calculating the hadronic parity violating matrix element presented in \cite{Kurth:2015}. Our method is particularly successful in the $^3P_2$ channel, even with the neglect of partial wave mixing effects. This work represents a crucial step toward the study of few- and many-body nuclear systems, as well as multi-nucleon matrix elements.

\bibliographystyle{physrev} 
\bibliography{NN} 

\end{document}